\documentclass[12pt,preprint]{aastex}
\usepackage{graphicx}

\def\kms{\mbox{$\rm km~s^{-1}$}}
\def\sec{$^{\prime\prime}$~}

\def\arcsec{^{\prime\prime}~}
\begin{document}

\title{CO Detection and Millimeter Continuum Emission from Low Surface Brightness Galaxies}

\author{Mousumi Das\altaffilmark{1, 2}, Karen O'Neil\altaffilmark{3}, 
Stuart N. Vogel\altaffilmark{2}, Stacy McGaugh\altaffilmark{2}} 

\affil{\altaffilmark{1}Raman Research Institute, Sadashivanagar, Bangalore, India}
\affil{\altaffilmark{2}Department of Astronomy, University of Maryland,
College Park, MD 20742}
\affil{\altaffilmark{3}NRAO, PO Box 2, Green Bank, WV 24944}

\email{email : mousumi@rri.res.in}

\begin{abstract}

We present BIMA and IRAM CO(1--0) observations of seven low surface 
brightness (LSB) galaxies,
including three large spirals with faint disks but prominent bulges, and four
relatively small LSB galaxies with irregular disks.
The giant LSB galaxies are UGC~5709, UGC~6614 and F568-6 (Malin2).
The smaller LSB galaxies are NGC~5585, UGC~4115, UGC~5209 and 
F583-1. The galaxies were selected based on their 
relatively high metallicity and apparent signs of  
star formation in their disks. The BIMA maps suggested the presence of molecular
gas in 2 of the giant LSB galaxies,
F568-6 and UGC~6614. Using the 30m IRAM telescope we
detected CO (1--0) emission in the disks of both galaxies and in the nucleus 
of F568-6. The molecular gas in these galaxies is clearly offset
from the nucleus and definitely associated with the LSB 
disk. In addition we also detected a millimeter 
continuum source in the center of UGC~6614. When compared with VLA 1.5~GHz observations
of the galaxy, the emission was found to have a flat spectrum indicating that the 
millimeter continuum emission is most likely due to an active galactic nucleus (AGN)
in the galaxy.
Our results show that giant LSB spirals may contain significant
quantities of molecular gas in their disks and also harbor radio bright AGN 
in their centers.
 
\end{abstract}


\section{INTRODUCTION}

Low surface brightness (LSB) galaxies have faint stellar disks of central 
brightness less than 23 B~mag~arcsec$^{-2}$ (Impey \& Bothun 1997). 
Although they have diffuse stellar disks
they are extremely gas rich with large HI gas disks that extend well beyond their 
apparent stellar disks. The high gas fraction observed in these
galaxies combined with 
the overall low metallicity suggests that they have had much less star formation 
compared to regular high surface brightness (HSB) galaxies, probably because their
gas surface densities are well below the critical value required for star formation
(van der Hulst et al. 1993). They are thus considered to be less evolved compared 
to HSB galaxies. They also have very large mass to light ratios which indicate
that their disks are dominated by massive dark matter halos; the massive halo inhibits
the formation of disk instabilities such as bars and spiral perturbations, which 
further contributes to the 
overall low star formation rate in these galaxies. 

LSB galaxies span a wide range of Hubble type and size but can be roughly divided into 
two categories. The first category is made up of the very populous dwarf spiral and
dwarf irregular galaxies. The majority of LSB galaxies in our local universe are of 
this type. The second category comprises of late type spiral galaxies that are 
very large in size; some may have prominent bulges and discernable spiral arms. 
These so called
``LSB giants" such as Malin~1 and UGC~6614 are relatively rare. The origin and 
evolution of these faint galaxies is still not clear. 
Previous studies suggest that LSB galaxies are relatively
isolated compared to other types of galaxies and the lack of galaxy interaction has 
left them less evolved compared to their HSB counterparts. To obtain a 
better understanding of their evolution, we need to understand their star formation 
history. An important step in this direction is to determine the gas
distribution in their disks. 

Although LSB galaxies are rich in HI, molecular gas (H$_{2}$) has been detected in 
only a handful of such galaxies (O'Neil, Hofner \& Schinnerer 2000; Matthews \& Gao 
2001; O'Neil, Schinnerer \& Hofner 2003; O'Neil \& Schinnerer 2004; Matthews et al. 2005).  
The low detection rate is probably a combination of several factors such as a low dust
content, low metallicity and a low surface density of neutral gas 
(de Blok \& van der Hulst 1998) all of which impede gas cooling, molecule formation
and cloud condensation.
However, modest amounts of ongoing star formation have been observed in most late type 
galaxies (McGaugh \& Bothun 1994; McGaugh, Schombert \& Bothun 1995; 
Matthews \& Gallagher 1997)
which must mean that star formation can proceed even in
these adverse environments. Very little is known about the physics of star formation in such
low density and low metallicity environments. One possibility is that star formation
occurs in localized regions over the disk. Molecular gas is then patchy in distribution 
and hence difficult to detect. 

To address some of these issues we searched for molecular gas in seven LSB galaxies. Three of
the galaxies are LSB giants and the remaining four are smaller late type galaxies.
The parameters of the galaxies
are listed in Table~1.  All the 
galaxies in our sample have relatively high metallicities and show
signs of star formation in their disks. 
We followed up tentative BIMA detections of CO (1--0) emission with 
IRAM 30m single dish observations. CO (1--0) emission was detected
in two galaxies, F568-6 (Malin~2) and UGC~6614. The molecular gas masses were derived
from the IRAM fluxes; they represent gas masses for the detected regions only and not
for the entire galaxy. 
We also detected millimeter continuum emission from the nucleus of one galaxy in our
sample, UGC~6614. The emission was detected at a mean frequency of 
111.2~GHz  and is most likely due to an AGN
in the galaxy.

In the following sections we present our observations and discuss the implications of our
results. For F568-6 (Malin~2) which has a fairly high redshift (z=0.046), we adopt a
luminosity distance of $D_{L}=198.8$~Mpc ($H_{o}=71~kms^{-1}Mpc^{-1}$); this
leads to a scale of 0.92~kpc arcsec$^{-1}$. UGC~6614 has a much lower redshift and so we
adopt the angular distance of $D=89.5$~Mpc ($V_{sys}=6351~\kms$),
which leads to a distance scale of 0.43~kpc arcsec$^{-1}$.

\section{OBSERVATIONS}

\subsection{GALAXY SAMPLE}
 
Our sample of seven galaxies is listed in Table 1. To cover the range of galaxy types, we
included in our sample both the large LSB spirals and the smaller LSB galaxies that have  
irregular disks. Also, 
to improve the chances of molecular gas detection, we chose galaxies that had
either relatively high metallicities or showed signs of star formation. Three of the 
galaxies in the sample, UGC~5709,
UGC~6614 and F568-6, are giant LSB galaxies. They 
are large, late type spirals with diffuse stellar disks but extended HI gas disks.
Of these three galaxies, UGC~6614 and F568-6 represent the prototypical giant LSB galaxies
(Quillen \& Pickering 1997).
Both are relatively bright and nearly face on, have prominent 
bulges and relatively strong spiral arms
(McGaugh, Schombert \& Bothun 1995). The HI gas distribution and kinematics for both
galaxies have been observed by Pickering et al. (1997).
UGC~6614 has a ring like feature around the 
bulge (Figure 1) which is prominent even in $H\alpha$ (McGaugh et al. 1995). It has
perhaps the highest metallicity known for an LSB galaxy and is estimated as
log(O/H)=-3 to -2.84, which is close to solar in value (McGaugh 1994).
Its nucleus shows AGN activity at optical wavelengths (Schombert 1998) and appears as a 
bright core in  X-ray emission (XMM archival data). F568-6, like UGC~6614, has a massive 
bulge and spiral structure; in addition it has several bright HII regions distributed 
over its inner disk. It also has a relatively high metallicity of log(O/H)=-3.22 which
is basically solar in value (de Blok \& van der Hulst 1998).
The HI velocity field has high velocities at one position
near the center which is associated with star formaion and may be due to
the accretion of a dwarf (Pickering et al. 1997). This galaxy also shows signs
of AGN activity at optical wavelengths (Schombert 1998) and its core has been detected
in X-ray emission (XMM archival data). UGC~5709 has a less regular structure than the former
2 galaxies and is intermediate in surface brightness. Its metallicity is log(O/H)=-3 to -3.2
which is approximately solar (de Blok \& van der Hulst 1998). 
Like F568-6 it has HII regions in 
the disk which have relatively high metallicities (McGaugh et al. 1995).

The other 4 galaxies are either dwarf galaxies or irregular galaxies that also have
relatively high metallicities or signs of star formation in their disks. 
They are NGC~5585, UGC~4115, UGC~5209 and F583-1. Apart from
F583-1, all the other galaxies are members of groups of galaxies. NGC~5585 is a companion
to the bright barred galaxy M101 and has HII regions distributed over the disk. Both UGC~4115
and UGC~5209 are members of loose groups of dwarf galaxies and have irregular morphologies.
F583-1 is the only distant galaxy in our sample of dwarf galaxies. It has a metallicity of 
log(O/H)$\sim$ -4 which is substantially subsolar (de Blok \& van der Hulst 1998). 
It also has a fairly symmetric
HI morphology (de Blok, McGaugh \& van der Hulst 1996). 

\subsection{BIMA OBSERVATIONS}

We observed the sample of galaxies with the BIMA interferometer 
(Welch et al. 1996) in the D array
from September 2002 to August 2003. The observations were all done in single pointing mode
with a 100\sec field of view centered on the galaxy nucleus. 
The galaxies were observed in the CO(1--0) emission line,
with the line positioned in the upper side band for all 7 galaxies. 
UGC~5709, UGC~6614
and F568-6 have fairly significant redshifts;  
the remaining 4 galaxies have low redshifts.
We used nearby radio sources for phase calibration; these are 
listed in Table 2. For flux calibration we used a nearby planet (e.g. Mars) or a radio
source whose flux is frequently monitored (3C~273). The data were reduced using MIRIAD (Sault,
Teuben \& Wright 1995). We used CLEAN and natural weighting for all our maps. The typical 
beam size for our maps was 15\sec ( see Table 2 for details).

We searched through the channel maps of each galaxy in our sample for
signs of CO (1--0) line emission (with velocity resolution of $\sim$6~\kms).
We also constructed the velocity-integrated CO (1--0) emission map by taking
the zeroth moment of the CO (1--0) emission. Our data suggest that two 
galaxies, UGC~6614 and F568-6, have CO (1--0) emission originating from their
disks. However, the emission lines for both galaxies were narrow and
noisy. Thus these observations were only an
indication that molecular gas may be present in these galaxies.
We did not see any CO (1--0) emission in the remaining five galaxies. However,
our non-detections put upper limits for
the CO (1--0) flux one could expect from these galaxies. 

In addition to searching for CO (1--0) line emission in the BIMA data, we looked for 
millimeter continuum 
emission from all seven galaxies in the sample. To do this we flagged channels that 
might contain
CO (1--0) line emission; the data flagging is based on the velocity distribution 
of HI emission. We then recomputed the continuum channels from the narrow band dataset. 
Then we averaged the upper and lower side band data to obtain the combined dataset. We again
used CLEAN and natural weighting for deriving the continuum emission maps.

\subsection{IRAM CO OBSERVATIONS}

To confirm the possible detections by BIMA of molecular gas ($H_{2}$) in these two
galaxies we observed them with the
IRAM 30m single dish telescope at positions close to the BIMA tentative detections.
UGC~6614 and F568-6 were observed in the CO (1--0) and (2--1) emission lines using
the IRAM 30m telescope in October--November of 2004 as part of the IRAM 30m Pool
Observations.
Table~3 lists the adopted positions, the frequencies and 
the total time each position was observed.
The beams were centered at the positions given in the
table and are 22\sec at 110 GHz and 11\sec at 220 GHz. The positions for IRAM 
pointing were chosen based on the tentative BIMA CO (1--0) line detections 
described in Section 2.2. 
Pointing  and focus were checked every 2-3 hours, depending on the weather
conditions
and pointing was found to be within the telescope limits (better than $2^{\prime\prime}$).
For each source both transitions were observed simultaneously with both frequencies
and two (circular) polarizations.
The backend was set using the 1MHz filterbank for each polarization for the 1mm lines
and the 4MHz filterbank for each of the 3mm lines.
The observed resolution of the 1mm and 3mm lines were
3.3 \kms\ and 5.3 \kms, respectively.

For data reduction, the lines were smoothed over 
seven channels to 23 \kms\ and over five channels to 27 \kms\ resolution, 
for the J(1$-$0) and J(2$-$1) 
lines, respectively, using the boxcar smoothing algorithm.
The maximal beam throw of $240^{\prime\prime}$ was used for these observations.
The image side band rejection ratios were measured
to be $>30$dB for the $3\,$mm SIS receivers and
$>12$dB for the $1.3\,$mm SIS receivers.  The data were
calibrated using the standard chopper wheel technique 
and are reported in main beam brightness temperature T$_{MB}$.
Typical system temperatures during the observations
were 160--180K and 170--250K in the $3\,$mm and $1.3\,$mm band, respectively.
All data reduction was done using CLASS -- the Continuum and Line Analysis
Single-dish Software developed by the Observatoire de Grenoble and IRAM (Buisson 2002).

\section{RESULTS}

\subsection{CO (1--0) Emission from UGC~6614 and F568-6}

Our BIMA observations suggested the presence of molecular gas at 
one location in the disk of UGC~6614 (Figures 1 \& 2) and at 
two positions in the disk of F568-6 (Figure 3, 4 and 5). As described above 
we followed up these observations 
with IRAM single dish observations centered near the BIMA tentative detections.
We detected CO (1--0) emission from the inner disk of both UGC~6614 and F568-6.
We did not detect any emission 
from the remaining 5 galaxies; however, the BIMA maps yield upper limits to the 
expected molecular gas masses in
these galaxies. 
In the following paragraphs we discuss the results for each galaxy in more detail.

\noindent
{\bf UGC~6614} : BIMA observations of UGC~6614 indicated that there may be 
CO (1--0) line emission originating
from the disk of UGC~6614 at approximately $38\arcsec$ west of the nucleus 
(Figure 1). The emission is weak and so the intensity map does not reveal much 
about the molecular gas distribution; hence we did not overlay it on the optical
image of the galaxy shown in Figure 1. However, though the 
line is weak (S/N$\sim3.6$) (Figure 2a) the velocity of the line is similar to the 
HI line velocity at that point. In the HI map of Pickering et al. (1997) the
HI velocity is between 6400 to 6420~\kms; this matches well with the BIMA line. 
Figure 1 also shows the
IRAM pointing center which is offset from the galaxy center by $38\arcsec$ and the 
IRAM beam which has a FWHM size of $22\arcsec$. The IRAM observations  
detected a significant amount of CO (1--0) flux (Figure 2b); no emission in the 
CO (2--1) line was detected. The emission is clearly offset from the galaxy center.
Thus the BIMA observations suggest that there may be significant CO (1--0)
emission from the LSB disk, west of the nucleus
and near the spiral arm/ring about the bulge. There may be molecular gas in the nucleus but 
it is too weak to be detected with BIMA. The molecular gas in the disk
appears to be associated with a spiral arm and extended along it. The IRAM observations 
confirm the presence of gas in the disk.
The CO emission may be associated with star formation knots in the inner ring.
To derive the molecular gas mass from the single dish CO flux, we used the
standard conversion factor which gives the molecular mass as 
$M_{mol}=1.5\times10^{4}{D_{Mpc}}^{2}S_{CO}$~M$_{\odot}$, where $D_{Mpc}$ 
is the galaxy distance in Mpc and 
$S_{CO}$ is the velocity integrated flux in Jy~\kms (Strong et al. 1988; 
Scoville et al. 1987).  
The total molecular gas mass observed in the disk from this region is 
$\sim2.8\times10^{8}~M_{\odot}$ (Table 4). 

\noindent
{\bf F568-6} : BIMA observations of F568-6 suggested that there may 
be CO (1--0) emission originating 
from 2 positions in the inner disk of the galaxy.  One position was close
to the nucleus (F568-6B) and the other was clearly offset from 
the nucleus and close to a spiral 
feature in the disk
(F568-6A). Although the emission lines seen in the BIMA data were relatively weak, the 
velocities of the lines match observed HI velocities at both positions and hence appear to be
real. Figure 3 shows the BIMA CO intensity map superimposed on the R band image of the galaxy. 
The IRAM pointings were at two locations in the disk and are marked as A and B in the figure;
they are offset by $\sim35\arcsec$ and $\sim7\arcsec$ respectively from the galaxy center.
Also shown is the IRAM beam that has a FWHM of $22\arcsec$. 

The IRAM CO(1--0) emission from position A (Figure 4) has
a line velocity of approximately 13990~\kms. This matches the   
HI velocity observed at that position by Pickering et al. (1997) which lies between 13980 
to 14020~\kms. The corresponding BIMA line (Figure 4a) is at a velocity of $\sim14010$~\kms 
which is slightly offset from the center of the IRAM line (Figure 4b). This may be 
because the molecular gas is distributed as clumps over the beam rather than a large
gas cloud. Our BIMA observations are thus picking up emission from one bright spot 
whereas the IRAM beam is sensitive to emission from the entire cloud complex. 
The emission from A is clearly positioned away from the galaxy nucleus; it lies to the east 
of the galaxy center and appears to be associated with a 
tightly wound spiral arm emerging from the edge of the bulge.
The IRAM CO(1--0) emission from position B has a peak velocity of approximately 13910~\kms 
(Figure 5) and the HI velocity at that position is between 13900 to 13920~\kms 
(Pickering et al. 1997). This emission appears to be associated with the nuleus. For 
this position, the BIMA and IRAM line velocities are fairly well matched (Figure 5a and b). 

To derive the molecular gas masses we used the CO luminosity ${L_{CO}}^{'}$ i.e. 
$M_{mol}=\alpha\times{L_{CO}}^{'}$~M$_{\odot}$, where 
$\alpha=4~M_{\odot}~{(K~km~s^{-1}~pc^{2})}^{-1}$ (Strong et al. 1988; Scoville et al. 1987).
The total molecular gas mass derived from summing up the gas detected at both positions
A and B using the IRAM CO (1--0) flux is
$\sim2.6\times10^{9}~M_{\odot}$ (Table 4). The emission is
extended east of the galaxy center but the 
molecular gas may well be extended west of the nucleus as well (Figure 3). This is because
the correlator setup for this galaxy did not cover the full range of velocities
for which HI emission has been detected in this galaxy, since the lower edge of the correlator
band was set at 13,860~\kms.
Thus our present detection is only a lower limit to the molecular gas distribution
in F568-6 and is based only on the redshifted half of the galaxy. Hence the molecular gas 
may be much more extended in the inner disk of the galaxy than our 
present observations indicate. In the near future we plan to map the extended molecular gas 
distribution in this galaxy in much better detail. 

\noindent
{\bf Remaining Galaxies :} We did not detect any CO (1--0) emission from the remaining five
galaxies in our sample (i.e. UGC~5709, NGC~5585, UGC~4115, UGC~5209 and F583-1).
Of these galaxies one
is a giant LSB galaxy (UGC~5709) and the others are dwarf galaxies. 
However we have used the noise levels in the BIMA maps to derive upper estimates of 
the molecular gas masses in these galaxies (Table 4) using the approximation 
$S_{CO}=(noise)\times(\delta v)\times(N_{b})$ where $\delta v$ represents the maximum width
of the expected CO (1--0) emission line; $N_{b}$ is number of beams in the $44\arcsec$ BIMA 
primary beam
(i.e. $\frac{{44\arcsec}^2}{b_{1}\times b_{2}}$) and $b_{1}, b_{2}$ are the beam sizes.
For $\delta v$ we have used the width of the HI line ($W_{50}$) as an upper estimate. Upper
limits to the molecular gas masses is calculated from the flux using the previously 
mentioned formula $M_{mol}=1.5\times10^{4}{D_{Mpc}}^{2}S_{CO}$~M$_{\odot}$ (Table 4). 

\subsection{3~mm Continuum Emission from UGC~6614} 

We searched for 3mm continuum emission from all the galaxies in our sample. 
Of the  seven galaxies 
that we observed two have been detected in radio continuum at 1.4~GHz 
in the NRAO VLA Sky Survey
(NVSS) (Condon et al. 1998); they are UGC~6614 and F568-6. 
Both galaxies are also strong emitters in the VLA
FIRST Survey (Becker, White \& Helfand 1995) which is at a similar 
frequency but has a much higher 
resolution than NVSS ($\sim5$\sec). 

We detected only UGC~6614 in 3mm continuum emission using BIMA (Figure 6); the remaining
6 galaxies were not detected. Table 5 lists the continuum 3mm flux observed in UGC~6614 and 
upper limits for the remaining galaxies. In UGC~6614 the continuum emission is detected at 
a mean frequency of 111.2~GHz and the peak flux at this 
frequency is 4.9~mJy/beam. The noise in the map is 1.2~mJy/beam which means that the 
detection has a signal to noise ratio (S/N) of approximately 4. The beam is 
$\sim16$\sec. The position of the peak in the continuum source is 
$11^{h}39^{m}14^{s}.8$, $17^{h}08^{m}37^{s}.5$; the error is $\pm4\arcsec$. Within error
limits the continuum emission peak is definitely coincident with the 
2MASS location of the galaxy center (Table 1). 
Although the S/N of the detection is not high, 
the fact that the continuum source is located at the
galaxy nucleus raises the significance of the detection.

We calculated the spectral index of the continuum emission for UGC~6614 and upper limits for
F568-6 using data from the FIRST 1.4~GHz VLA survey. We
convolved the FIRST 1.4~GHz maps of UGC~6614 and F568-6  
to the same resolution as the BIMA maps and measured the peak flux. Table 5 lists the spectral
index for UGC~6614  and upper limits for F568-6. The spectrum is flat for UGC~6614 ($\alpha\sim0$ 
where $f_{\nu}\propto\nu^{\alpha}$) and is likely declining at millimeter wavelengths for F568-6.
As both galaxies host AGN and have little star formation activity, the 
radio continuum emission is probably non-thermal in nature. However, for UGC~6614, we confirmed
this by using  
the $H\alpha$ flux in the galaxy (de Blok \& van der Hulst 1998) to derive the
star formation rate (SFR) and infrared luminosity (Kennicutt 1983; Kennicutt 1998). The 
expected 1.4~GHz radio
continuum flux from the $H\alpha$ emission is only 0.15~mJy. This is much less than the 
actual continuum flux
measured in the 1.4~GHz map of UGC~6614 which is 5.67~mJy. Thus the BIMA 3mm continuum
emission from UGC~6614, is  non-thermal in nature and
due to an AGN in the galaxy. As mentioned earlier, AGN activity in UGC~6614 has been detected at  
optical wavelengths (Schombert 1998) and also in X-ray
emission (XMM archive). 

\section{DISCUSSION}

We have detected CO (1--0) line emission from the disks of two giant LSB galaxies, F568-6 and 
UGC~6614. We have also detected and mapped the millimeter continuum source 
in the nucleus of UGC~6614. In the
following paragraphs we discuss the implications of our findings.

\subsection{Molecular Gas Detection}

As mentioned in Section 1, very few LSB galaxies have been detected in CO emission. Of the handful
of galaxies in which molecular gas has been detected some are giant, bulge dominated spirals with
large disk rotation velocities (O'Neil et al. 2000, 2003) while others are more low mass, disk
dominated systems (Matthews \& Gao 2001, Matthews et al. 2005). However, in nearly all previous detections
the molecular gas has been found to be centrally concentrated and associated with the galaxy 
bulge or nucleus. The exception being the single dish observations of Matthews \& Gao (2001) 
which show CO emission originating from off center positions for two galaxies in their sample.
Our detections of CO emission from UGC~6614 and F568-6 are however  
unique because both detections have interferometric as well as single dish observations of the molecular gas
distribution that clearly indicate that they are associated with the disk and not just the galaxy center. 
This is important as it shows that molecular gas can form and exist in the disks of LSB  
galaxies as well as in their nuclear regions.

Previous studies of LSB galaxies suggest that the low detection rate of molecular gas is a result of
a combination of low metallicity and low gas surface density in the disk (van der Hulst et al. 1993;
de Blok et al. 1996; Mihos, Spaans \& McGaugh 1999). In addition LSB galaxies have massive dark halos
that may prevent global disk instabilities from forming even when they interact with other galaxies
(Mihos, McGaugh and de Blok 1997); this reduces the formation of shocks that form 
dense clouds and lead to massive star formation in galaxy disks (Gerritsen \& de Blok 1999).  
In LSB disks the molecular gas 
may not form due to such large scale instabilities. Instead it may form due to  
local instabilities which will result in isolated clumps of molecular gas and localized regions of 
modest star formation. Signatures of such localized star formation has  
been observed in LSB galaxies (Schombert et al. 1990; McGaugh et al. 1995) and our observations
of molecular gas in the disks of both UGC~6614 and F568-6 further support this idea.
Thus LSB
galaxies may not be as devoid of molecular gas as thought earlier; the distribution may
just be localized to isolated regions over the disk
and hence difficult to detect.

Another possibilty that is hard to exclude is that given the physical conditions of the ISM
in LSB galaxies, the CO to H$_{2}$ conversion factor is different from the standard value
(Mihos et al. 1999; Gerritsen \& de Blok 1999). This might allow for substantially more molecular 
gas to be present than we infer. Bearing this caveat in mind, our observations nevertheless seem 
to suggest a picture in which molecular gas is rare in LSB galaxies but not completely absent.
Such a distribution of molecular gas supports the observed, low intensity star formation in 
isolated regions that is seen in these galaxies. 
 
\subsection{Continuum Emission}

Figure 6 shows the BIMA map of the continuum source in UGC~6614. The central peak coincides 
with the 2MASS center of the galaxy; the offset between the two is less than an arcsecond.  
The emission is probably non-thermal in nature and due to an AGN in the galaxy. 
Several giant LSB galaxies that have prominent bulges have been
found to host AGN and show weak Seyfert activity. AGN activity in LSB galaxies 
has been detected in optical emission lines (Schombert 1998;
Sprayberry et al 1995). The nuclear activity is due to the accretion of
mass onto a supermassive black hole in the galactic center. It is now
widely accepted that most galaxies with a well defined bulge contain a
black hole (Magorrian et al. 1998) and that the bulge mass correlates well
with the black hole mass (Ferrarese et al. 2001). Giant LSB galaxies have
both prominent bulges and copious amounts of neutral hydrogen (HI) gas
in their inner disks. Hence it is not suprising that these galaxies show
signs of AGN activity. However it is not clear how the gas is transported to the nucleus 
since gas fueling processes such as bars are not frequently found in these
galaxies (Mihos et al. 1997). It could be that since the AGN activity is 
fairly weak in LSB galaxies, the gas torquing produced by the disk spiral
arms is enough to funnel gas into the nuclear regions and 
fuel the AGN. 

Very little is known about the continuum emission from
LSB galaxies. A handful appear bright in the NVSS and FIRST surveys which are at 1.4~GHz, but
none have been studied at millimeter wavelengths (e.g. 100~GHz). The continuum source at 111~GHz
in UGC~6614 is the first such detection at millimeter wavelenghths. Suprisingly F568-6 is
not bright at millimeters wavelengths although its core is bright in the 1.4~GHz continuum maps 
(NVSS, FIRST). This implies that the synchroton  spectrum of F568-6 turns over from a flat
spectrum to a steeply falling spectrum before millimeter wavelengths, whereas in UGC~6614
it remains constant. A similar difference in turnover frequencies is observed in the nearby
Seyfert galaxies NGC~1068 and NGC~3147 (Krips et al. 2005) and is probably due to the
different magnetic field strengths in their AGN (e.g. Krolik 1999). Flat spectrum radio 
galaxies also show a variation in turnover frequencies between centimeter to 
millimeter wavelenghths (Bloom et al. 1999) but not many sources are found to be bright at
millimeters. Thus UGC~6614 is unique with respect to its nuclear continuum source which is radio
bright even at millimeter wavelengths.

\section{CONCLUSIONS}

We have searched for molecular gas in a sample of seven LSB galaxies using the BIMA interferometer.
Molecular gas was detected in the disks of two galaxies, 
UGC~6614 and F568-6, using the IRAM 30m single dish 
telescope. Both galaxies have prominent bulges and large, low surface brightness
disks. Our results indicate that molecular gas may be present in both the 
disks of LSB galaxies as
well as their nuclei but the distribution is localized over isolated regions 
and thus difficult to detect in unbiased single dish observations. Overall it appears that
molecular gas is rare in these galaxies but not completely absent.
We have also detected millimeter continuum emission from the nucleus  
of one of these galaxies, UGC~6614. Our main results are summarised below.

1. CO (1--0) emission was detected in UGC~6614 at approximately 17~kpc (38\sec)
west of the nucleus. It is associated with a spiral feature in the disk.
The mass of molecular gas detected from this location is 
$\sim2.8\times10^{8}~M_{\odot}$. The BIMA observations did not detect any
molecular gas in the nuclear region, probably because the gas surface density 
averaged over the BIMA beam falls below the detection limits
of the observation. 

2. CO (1--0) emission in F568-6 was detected at two positions in the inner disk.
One position is close to the nucleus and the other is approximately 
28~kpc (30\sec) east of the
nucleus. The BIMA intensity maps indicate that the CO (1--0) emission is distributed 
about the nucleus and extends into the inner disk. The mass of molecular gas associated
with the detection is $\sim2.6\times10^{9}~M_{\odot}$. 
  
3. We have also detected a millimeter continuum source in UGC~6614 at the center of the galaxy.
Comparing it with VLA FIRST 1.5~GHz maps, we find that it has a flat spectrum between ~ 1 to 
110~GHz. The continuum emission is due to the AGN in the galaxy which has been detected at
optical and X-ray wavelengths as well. Millimeter continuum emission from Seyferts or
radio loud galaxies is rare and so UGC~6614 is fairly unique in its AGN activity compared to 
most spiral galaxies. 

\acknowledgments

The authors would like to thank Alice Quillen for providing the R band images 
of UGC~6614 and F568-6.
Observations with the BIMA millimeter-wave
array are partially supported by NSF AST-0228974.  
We are grateful to the members of the IRAM staff for their help in the
observations. This research has
made use of the NASA/ IPAC Infrared Science Archive, which is operated
by the Jet Propulsion Laboratory, California Institute of Technology,
under contract with the National Aeronautics and Space Administration. 
We also acknowledge the usage of the HyperLeda database (http://leda.univ-lyon1.fr).

\newpage
{\scriptsize
\begin{deluxetable}{lcccccc}
\tablenum{1}
\tablewidth{0pt}
\tablecaption{Galaxy Sample}
\tablehead{
\colhead{Galaxy} & \colhead{Galaxy} & \colhead{Velocity} & \colhead{Galaxy Position} & \colhead{$D_{25}$} & \colhead{inclination} \\ 
\colhead{Name} & \colhead{Type} & \colhead{\kms} & \colhead{RA, $\delta$ (J2000)} & \colhead{$\arcsec$} & \colhead{degrees}
}
\startdata
UGC~5709 & Sd        & 6206 & $10^{h}31^{m}16^{s}.2$, $+19^{\circ}22^{\prime}59^{\prime\prime}$ & 80.9 & 54.6 \\
UGC~6614 & (R)SA(r)a & 6351 & $11^{h}39^{m}14^{s}.8$, $+17^{\circ}08^{\prime}37^{\prime\prime}$ & 99.6 & 29.9 \\
F568-6   & Sd/p      & 13820 & $10^{h}39^{m}52^{s}.5$, $+20^{\circ}50^{\prime}49^{\prime\prime}$ & ... & 38.0  \\
NGC~5585 & SAB(s)d   & 305  & $14^{h}19^{m}48^{s}.2$, $+56^{\circ}43^{\prime}45^{\prime\prime}$ & 345.3 & 53.2 \\
UGC~4115 & IAm       & 338  & $07^{h}57^{m}01^{s}.8$, $+14^{\circ}23^{\prime}27^{\prime\prime}$ & 109.2 & 67.1 \\
UGC~5209 & Im        & 538  & $09^{h}45^{m}04^{s}.2$, $+32^{\circ}14^{\prime}18^{\prime\prime}$ & 54.7 & 0.0 \\
F583-1   & Sm/Irr    & 2264 & $15^{h}57^{m}27^{s}.5$, $+20^{\circ}39^{\prime}58^{\prime\prime}$ & 42.0 & 63.0 \\ 

\enddata
\begin{flushleft}
(a)~The $D_{25}$ diameters for all the galaxies is from the RC3 catalogue, except for F583-1 which is from 
de Blok, McGaugh \& van der Hulst (1997).\\
(b)~The inclinations for all the galaxies except F568-6 and F583-1 is from Hyperleda catalogue 
(Paturel et al. 2003). For F568-6, the
inclination is from Pickering et al. (1997) and for F583-1 the inclination is obtained from de Blok, McGaugh \&
van der Hulst (1997).
\end{flushleft}
\end{deluxetable}

\newpage
{\scriptsize
\begin{deluxetable}{lcccccc}
\tablenum{2}
\tablewidth{0pt}
\tablecaption{BIMA Observations}
\tablehead{
\colhead{Galaxy} & \colhead{CO(1--0)}         & \colhead{Phase}      & \colhead{Flux} & \colhead{Beam} & \colhead{Number}\\
\colhead{Name  } & \colhead{Frequency (GHz)}  & \colhead{Calibrator} & \colhead{Calibrator} & \colhead{Size} & \colhead{of tracks}
}
\startdata
UGC~5709 & 112.93 & 1058+015 & 3C~273 & $18.5\arcsec\times14.5\arcsec$ & 3 \\
UGC~6614 & 112.88 & 1118+125 & Mars   & $17.3\arcsec\times15.0\arcsec$ & 3 \\
F568-6   & 110.19 & 1058+015 & 3C~273 & $18.4\arcsec\times16.5\arcsec$ & 6 \\
NGC~5585 & 115.15 & 1419+543 & Mars   & $17.3\arcsec\times13.9\arcsec$ & 3 \\
UGC~4115 & 115.14 & 0750+125 & Mars   & $17.3\arcsec\times14.6\arcsec$ & 1 \\
UGC~5209 & 115.07 & 0927+390 & W3OH   & $16.9\arcsec\times13.8\arcsec$ & 2 \\
F583-1   & 114.41 & 1540+147 & Mars   & $16.4\arcsec\times15.7\arcsec$ & 1 \\

\enddata
\end{deluxetable}

\newpage
{\scriptsize
\begin{deluxetable}{lccccc}
\tablenum{3}
\tablewidth{0pt}
\tablecaption{IRAM Observations}
\tablehead{
\colhead{Galaxy} & \colhead{RA}& \colhead{DEC} & \colhead{CO(1--0)}        & \colhead{CO(2--1)} & \colhead{Time}\\
\colhead{Name  } & \colhead{}  & \colhead{ }   & \colhead{Frequency (GHz)} & \colhead{Frequency (GHz)}& \colhead{in mins}
}
\startdata
F568-6a & $10^{h}39^{m}55^{s}$ & $20^{h}50^{m}54^{s}$ & 110.20559 & 220.40696 & 134 \\
F568-6b & $10^{h}39^{m}53^{s}$ & $20^{h}50^{m}50^{s}$ & 110.31817 & 220.63212 & 240 \\
UGC~6614 & $11^{h}39^{m}12^{s}$ & $17^{h}08^{m}31^{s}$ & 112.89500 & 225.78569 & 184 \\

\enddata
\label{iramobs}
\end{deluxetable}

\newpage
{\scriptsize
\begin{deluxetable}{lccccccc}
\tablenum{4}
\tablewidth{0pt}
\tablecaption{CO Fluxes and Molecular Gas Masses}
\tablehead{
\colhead{Galaxy} & \colhead{BIMA Peak} & \colhead{Noise} & \colhead{Channel} & \colhead{HI Width} & \colhead{IRAM} &
\colhead{Noise} & \colhead{Molecular}\\
\colhead{Name} & \colhead{Flux}  & \colhead{} & \colhead{Width} & \colhead{W$_{50}$} & \colhead{Flux} & \colhead{} & \colhead{Gas Mass$^a$}\\
\colhead{} & \colhead{(Jy~beam$^{-1}$)} & \colhead{} &\colhead{(\kms)} & \colhead{(\kms)}& \colhead{(Jy~\kms)} & \colhead{} & \colhead{(M$_\odot$)}
}
\startdata
UGC~6614    &   0.060  & 0.017 & 6.1  & 242  & 2.30   &    0.006   &  $2.8\times10^{8}$ \\
F568-6~A    &   0.043  & 0.013 & 6.1  & .... & 1.86   &    0.005   &  $0.7\times10^{9}$ \\    
F568-6~B    &   0.057  & 0.013 & 6.1  & .... & 5.24   &    0.005   &  $1.9\times10^{9}$ \\
\hline
UGC~4115    &   ...    & 0.047 & 6.1  & 83   & ....   &     ....   &  $<~1.1\times10^{7}$ \\  
UGC~5209    &   ...    & 0.007 & 6.1  & 49   & ....   &     ....   &  $<~2.4\times10^{6}$ \\  
UGC~5585    &   ...    & 0.039 & 6.1  & 146  & ....   &     ....   &  $<~1.3\times10^{7}$ \\
UGC~5709    &   ...    & 0.022 & 6.1  & 247  & ....   &     ....   & $<~4.6\times10^{9}$  \\
F583-1      &   ...    & 0.056 & 6.1  & 174  & ....   &     ....   & $<~1.1\times10^{9}$  \\ 

\enddata

\begin{flushleft}
(a)~Molecular gas masses were derived from IRAM fluxes; the values represent gas masses 
for the detected regions only and not for the entire galaxy. The upper limits for the 
non-detections in BIMA were derived using the noise levels in the corresponding channel 
maps and the width of the HI line ($W_{50}$).
For F583-1, twice the HI rotation curve peak was used (de Blok, McGaugh \& Rubin 2001)
for width of the HI line instead of $W_{50}$.\\
\end{flushleft}
\label{fluxobs}
\end{deluxetable}

\newpage
{\scriptsize
\begin{deluxetable}{lcccccc}
\tablenum{5}
\tablewidth{0pt}
\tablecaption{Continuum Fluxes and Spectral Indices}
\tablehead{
\colhead{Galaxy} & \colhead{BIMA Continuum} & \colhead{Frequency} & \colhead{FIRST Continuum} & \colhead{Frequency} & \colhead{Spectral Index} \\
\colhead{Name} & \colhead{Flux (mJy/b)~$^a$} & \colhead{GHz}      & \colhead{Flux (mJy/b)~$^b$} & \colhead{GHz}       & \colhead{ }
}
\startdata
UGC~6614         &  4.9      &  111.25  &  5.5  & 1.4  & -0.03       \\
F568-6           &  $<~2.4$  &  108.4   &  5.4  & 1.4  & $<~-0.19$   \\
\hline
UGC~4115         &  $<~8.4$  &  113.5   &  ..   &  ..  & ..            \\ 
UGC~5209         &  $<~1.2$  &  113.5   &  ..   &  ..  & ..             \\
NGC~5585         &  $<~6.3$  &  113.5   &  ..   &  ..  & ..             \\
UGC~5709         &  $<~4.5$  &  111.3   &  ..   &  ..  & ..             \\
F583-1           &  $<~10.5$  &  112.8   &  ..   &  ..  & ..             \\ 
\enddata
\begin{flushleft}
(a)~The upper limits for the non-detections of BIMA continuum fluxes are given as three 
times the noise levels in the corresponding maps.\\
(b)~The FIRST map is smoothed to the same resolution of the BIMA map before estimating the 
continuum flux.
\end{flushleft}
\label{cont_flux}
\end{deluxetable}

\newpage
\centerline{\bf FIGURE CAPTIONS }

\noindent 
Figure 1. The R band optical image of UGC~6614; the location of the center of 
the IRAM beam is indicated with the arrow labeled A. The telescope pointing center
is well away from the center of the galaxy. The IRAM beam size is 22\sec and
the 2MASS center of the galaxy is marked with a filled triangle.  
                                                                                                 
\noindent
Figure 2. Panel showing the BIMA and IRAM CO (1--0) emission lines observed in
UGC~6614. The topmost panel is a CO spectrum observed by BIMA from the nuclear region; 
no CO (1--0) emission is seen here. The middle panel shows the BIMA CO (1--0) spectrum 
observed from location A that lies west of the nucleus and in the LSB disk.
Both spectra have been smoothed to a velocity resolution of
6~\kms. The lower panel is the IRAM single
dish CO (1--0) spectrum smoothed to a velocity resolution of 27~\kms. The
systemic velocity of the galaxy is marked on the x axis with an
arrow. The dashed lines show the approximate width of the single dish emission line. 

\noindent
Figure 3. Contours of BIMA C0 (1--0) line emission overlaid on the R band optical
image of F568-6 (Malin~2). The intensity contours in white are at 1.5, 2, 2.5, 
3 and 3.5 
times the rms noise level and the contours in black are at -1.5 and -2.0 times 
the rms noise level. The 
beam is $\sim17$\sec. The pointing centers of the 22\sec IRAM beams for 
the two C0 (1--0) emission detections are indicated with the arrows labeled A and B.
The 2MASS center of the galaxy is marked with a filled triangle.

\noindent
Figure 4. Panel showing the BIMA and IRAM CO (1--0) emission lines observed in
F568-6 from the location marked A in Figure 3 (i.e. approximately 35\sec west
of the nucleus). The
upper panel shows the BIMA CO (1--0) spectrum smoothed to a velocity resolution of
6~\kms; the lower panel shows the IRAM single
dish CO (1--0) spectrum smoothed to a velocity resolution of 27~\kms. The
systemic velocity of the galaxy is marked on the x axis with an
arrow. The dashed lines show the approximate width of the single dish emission line.

\noindent
Figure 5. Panel showing the BIMA and IRAM CO (1--0) emission lines observed in
F568-6 from the location marked B in Figure 3 (i.e. close to the nucleus). The
upper panel shows the BIMA CO (1--0) spectrum smoothed to a velocity resolution of
6~\kms; the lower panel shows the IRAM single
dish CO (1--0) spectrum smoothed to a velocity resolution of 27~\kms. The
systemic velocity of the galaxy is marked on the x axis with an
arrow. The dashed lines show the approximate width of the single dish emission line.

\noindent
Figure 6. BIMA continuum emission map of UGC~6614 at 2.7mm.
The emission is averaged over the upper and lower sidebands, with a mean frequency
of 111.2~Ghz. The emission is centered on the 2MASS near-infrared position of the 
nucleus. The beam is shown on the lower left. The emission is contoured at 
-2, -2.5, 2, 2.5,
3, 3.5 and 4 times the noise level. The noise level is 1.2~mJy/beam where the beam is
$17.4\times15$\sec. The peak emission is 4.9~mJy/beam. 

\newpage

\begin{references}
\reference {} Becker, R.H.; White, R.L.; Helfand, D.J. 1995, ApJ, 450, 559
\reference {} Bloom, S.D. et al. 1999, ApJS, 122, 1 
\reference {} Buisson, G. et al. 2002, CLASS Continuum and Line Analysis System Handbook
online at http://iram.fr/GS/class/class/html 
\reference {} Condon, J.J.; Cotton, W.D.; Greisen, E.W., Yin, Q.F., Perley, R.A., Taylor, G.B.
\& Broderick, J. 1998, AJ, 115, 1693
\reference {} Cote, S., Carignan, C., Sancisi, R., 1991, AJ, 102, 904
\reference {} de Blok, W. J. G., McGaugh, S.S. \& van der Hulst, J.M. 1996, MNRAS, 283, 18 
\reference {} de Blok, W. J. G. \& van der Hulst, J. M. 1998, A\&A, 336, 49
\reference {} de Blok, W. J. G.; McGaugh, Stacy S.; Rubin, Vera C. 2001, AJ, 122, 2396
\reference {} Ferrarese, L. et al. 2001, ApJ, 555, L79 
\reference {} Gerritsen, J.P.E.  \& de Blok, W.J.G. 1999, A\&A, 342, 655
\reference {} Hoffman, G. Lyle; Salpeter, E. E.; Farhat, B.; Roos, T.; Williams, H.; Helou, G. 
1996, ApJS, 105, 269 
\reference {} Impey, C. \& Bothun, G. 1997, ARA\&A, 35, 267 
\reference {} Karachentsev, I.D., Karachentseva, V.E., Huchtmeier, W.K. 2001, A\&A, 366, 428
\reference {} Kennicutt, R.C. 1983, ApJ, 272, 54
\reference {} Kennicutt, R.C. 1998, ARA\&A, 36, 189
\reference {} Krips, M. et al. 2005, astro-ph/0509825
\reference {} Krolik, J.H. 1999, Active Galactic Nuclei, Princeton University Press
\reference {} Magorrian, J. et al. 1998, AJ, 115, 2285 
\reference {} Matthews, L.D. \& Gallagher, J.S. 1997, AJ, 114, 1899  
\reference {} Matthews, L.D. \& Gao, Y. 2001, ApJ 549, L191
\reference {} Matthews, L.D.; Gao, Y.; Uson, J.M.; Combes, F. 2005, AJ, 129, 1849
\reference {} McGaugh, S.S. 1994, ApJ, 426, 135
\reference {} McGaugh, S.S. \& Bothun, G.D. 1994; AJ, 107, 530 
\reference {} McGaugh, S.S.; Schombert, J.M. \& Bothun, G.D. 1995, AJ, 109, 2019 
\reference {} Mihos, J.C.; McGaugh, S.S. \& de Blok, W. J. G. 1997, ApJL, 477, L79 
\reference {} Mihos, J.C.; Spaans, M. \& McGaugh, S.S. 1999, ApJ, 515, 89
\reference {} O'Neil, K.; Hofner, P. \& Schinnerer, E. 2000, ApJ, 545, L102
\reference {} O'Neil, K.; Schinnerer, E.; \& Hofner, P. 2003, ApJ, 588, 230 
\reference {} O'Neil, K. \& Schinnerer, E. 2004, ApJL 615, 109
\reference {} Quillen, A. C.; Pickering, T. E. 1997, AJ, 113, 2075 
\reference {} Paturel, G.; Theureau, G.; Bottinelli, L.; Gouguenheim, L.;
Coudreau-Durand, N.; Hallet, N.; Petit, C. 2003, A\&A, 412, 45
\reference {} Pickering, T. E.; Impey, C. D.; van Gorkom, J. H.; Bothun, G. D. 1997, AJ, 114, 1858
\reference {} Sault, R. J.; Teuben, P. J.; Wright, M. C. H. 1995, ADASS, ASP Conference Series, Vol. 77, eds. R.A. Shaw, H.E. Payne, and J.J.E. Hayes, p. 433.
\reference {} Schombert, J.M.; Bothun, G.D.; Impey, C. D. \& Mundy, L.G. 1990, AJ, 100, 1523
\reference {} Schombert, J.M. 1998; AJ, 116, 1650
\reference {} Scoville, N. Z.; Sanders, D. B. 1987, in Interstellar processes, ed. D. Hollenbach \& H. Thronson (Dordrecht, D. Reidel), p.21
\reference {} Sprayberry, D; Impey, C.D.; Bothun, G.D. \& Irwin, M.J. 1995, AJ, 109, 558  
\reference {} Strong, A. W. et al. 1988, A\&A, 207, 1
\reference {} van der Hulst, J.M.; Skillman, E.D.; Smith, T.R.; Bothun, G.D.; 
McGaugh, S.S. \& de Blok, W.J.G. 1993 AJ, 106, 548
\reference {} Welch, W.J. et al. 1996, PASP, 108, 93


\end{references}
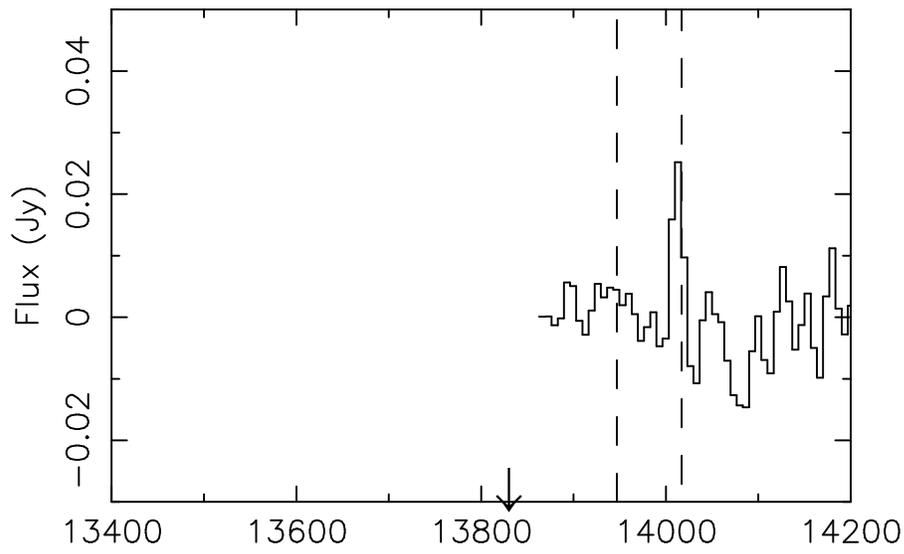
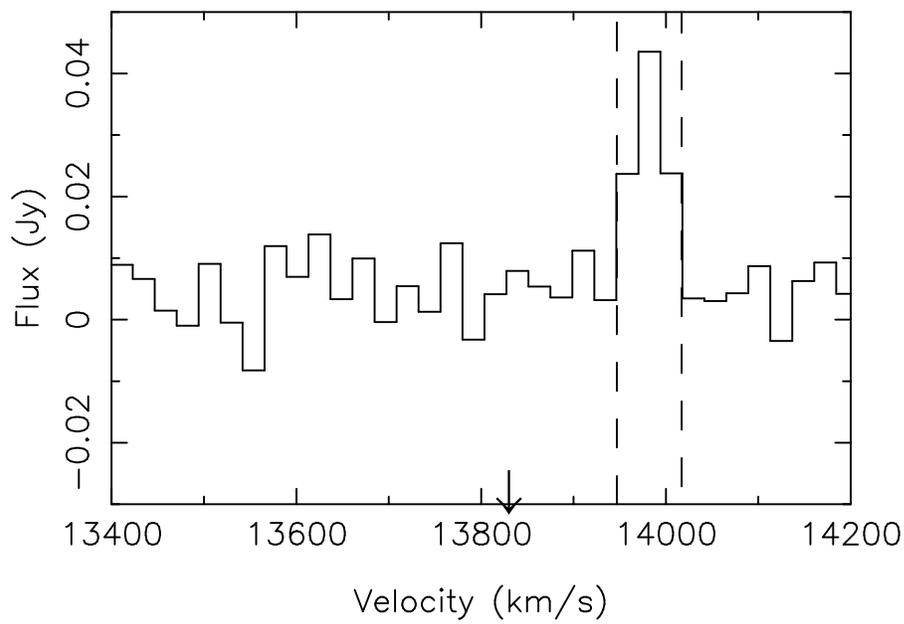
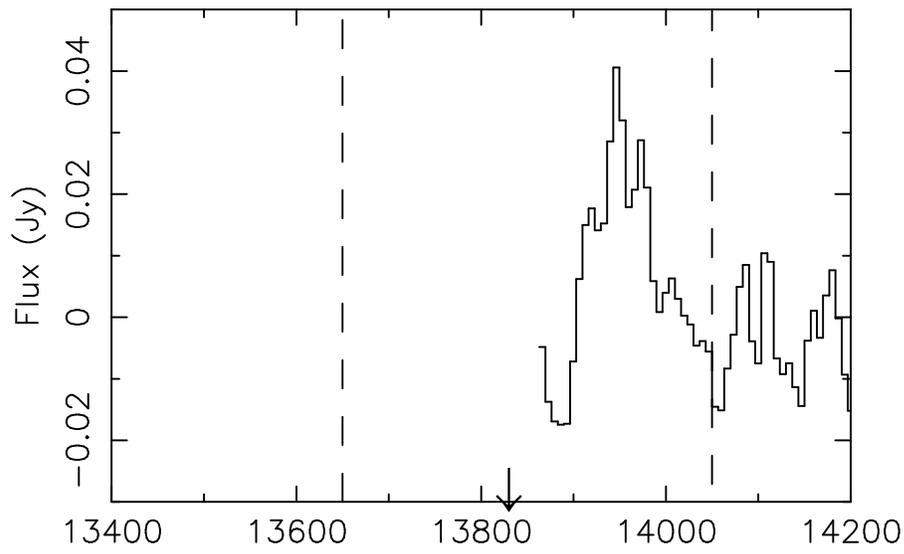
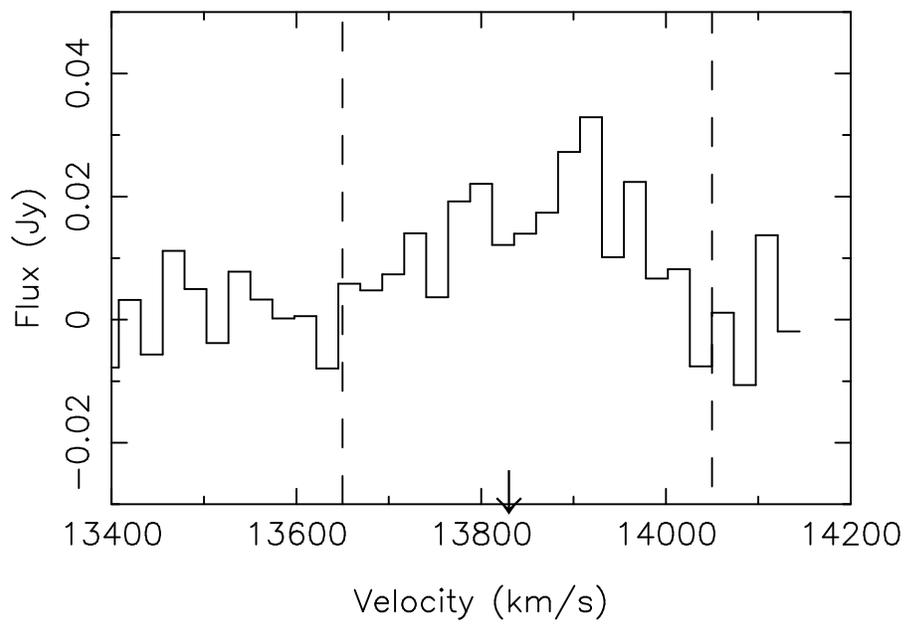
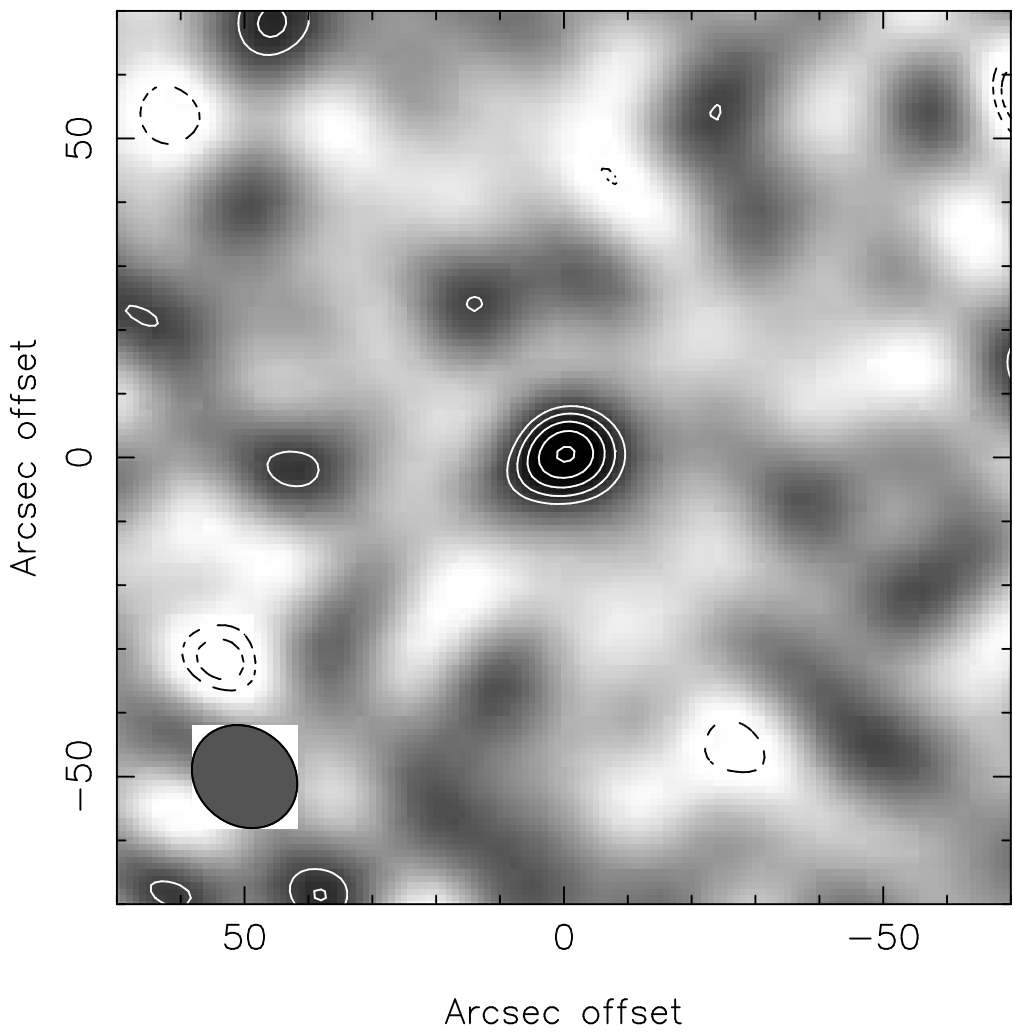
\end{document}